\documentclass[eqsecnum,preprint]{aastex}

\slugcomment{YITP-99-54,~ astro-ph/9909183}

\newcommand{\om}{\Omega_0}
\newcommand{\beq}{\begin{equation}}
\newcommand{\eeq}{\end{equation}}
\newcommand{\bea}{\begin{eqnarray}}
\newcommand{\eea}{\end{eqnarray}}

\newcommand{\lmu}{_{\mu}}
\newcommand{\ijl}{_{ij}}
\newcommand{\ij}{^{i}_{j}}
\newcommand{\tv}{\tilde v}
\newcommand{\etal}{{et al.\ }}
\newcommand{\ptp}{{Prog. Theor. Phys.\ }}
\newcommand{\PR}{{Phys. Rev.\ }}

\shorttitle{Voids in McVittie Spacetime}
\shortauthors{Sakai, Haines}

\begin{document}

\title{Peculiar Velocities of Nonlinear Structure:\\
Voids in McVittie Spacetime}

\affil{Nobuyuki Sakai
\thanks{Electronic address: sakai@yukawa.kyoto-u.ac.jp}}
\affil{Yukawa Institute for Theoretical Physics, Kyoto University,
Kyoto 606-8502, Japan}
%\email{sakai@yukawa.kyoto-u.ac.jp}
\and
\affil{Paul Haines}
\affil{Department of Physics and Astronomy, Dartmouth College,
Hanover NH 03755, USA}

\begin{abstract}
As a study of peculiar velocities of nonlinear structure, 
we analyze the model of a relativistic thin-shell void in the expanding
universe.
(1) Adopting McVittie (MV) spacetime as a background universe, 
we investigate the dynamics of an uncompensated void with negative MV mass. 
Although the motion itself is quite different from that of a 
compensated void, as shown by \citet{hai93}, 
the present peculiar velocities are not affected by MV mass.
(2) We discuss how precisely the formula in the linear perturbation 
theory applies to nonlinear relativistic voids, using the results in 
(1) as well as the previous results for the homogeneous background
\citep{SMS93}.
(3) We re-examine the effect of the cosmic microwave background radiation.
Contrary to the results of
\citet{pim86,pim88}, we find that the effect is negligible. We show that their
results are due to inappropriate initial conditions.
Our results (1)-(3) suggest that the formula in the linear 
perturbation theory is approximately valid even for nonlinear voids.

%\vskip1cm \noindent
%PACS numbers: 98.65.Dx, 04.25.-g \\
\end{abstract}


~ \newpage
%%%%%%%%%%%%%%%%%%%%%%%%%%%%%%%%%%%%%%%
\section{Introduction}
%%%%%%%%%%%%%%%%%%%%%%%%%%%%%%%%%%%%%%%

Measurements of large-scale peculiar velocities can provide a constraint 
on the universe model (see, e.g. \citet{zeh99}). In contrast to other
cosmological 
tests, they give a constraint on the density parameter $\om$ alone,
almost independent of the cosmological constant $\lambda_0$ \citep{car92}.

Although the relation between the peculiar velocity 
and the density parameter $\om$ is usually given in the linear perturbation 
theory (LPT) \citep{pee76},  the observed universe has nonlinear density 
profiles. In fact, a network of nonlinear voids filling the entire 
universe has been suggested by redshift surveys such as the CfA2 \citep{gel89} 
and 
the SSRS2 \citep{da94}.  Moreover, using such redshift surveys, the 
description of a void-filling universe was confirmed quantitatively 
\citep{el97,EPD96,EPD97}.
The relation between $\om$ and peculiar
velocities inside underdense regions suggests $\om\leq 0.3$ can be
ruled out at the $2.4$-sigma level \citep{dek94}. 

It is therefore important to investigate peculiar velocities of 
nonlinear void structure. Here we consider the model of a relativistic 
thin-shell void.
The expansion law of relativistic voids was investigated originally by 
\citet{mae83a,mae83b}, developing the metric junction method proposed 
by \citet{isr66}. They found analytically that, in the flat
universe, the shell radius $R$ expands asymptotically as
$R \propto t^{(15+\sqrt{17})/24}\approx t^{0.797}$ \citep{mae83a}.
For the self-similar void model, on the other hand, \citet{bert85}
obtained the solution with $R\propto t^{0.8}$. The difference 
of the two results is so small that we do not have to care which model 
is better, which is physically determined by the radiative process in
the shell.
For other universe models, the motion of the shell was calculated numerically 
\citep{mae83b}: in the open universe, the shell expansion is eventually
frozen to the background expansion; on the other hand, in the closed universe, 
the shell expands much faster and its velocity finally approaches the speed of
light.
The relation between the peculiar velocity of the shell and
the universe model was later investigated systematically \citep{SMS93}.

Lake \& Pim extended the work of Maeda \& Sato so as to include a
mass inside a void \citep{lak85} and the cosmic microwave background (CMB) 
radiation \citep{pim86,pim88}. In particular, they claimed that the
inclusion of radiation has significant quantitative and qualitative effects on
the
evolution of the void. It was shown, for instance, that the asymptotic
behavior of the shell is $R\propto t$ in the flat universe if the CMB
radiation is included \citep{pim86}.
This result is in contrast to that for the vacuum 
void ($R \propto t^{0.8}$), and hence quite surprising. 

\citet{hai93}, on the other hand, included a mass 
{\it outside} a void by employing McVittie (1966, hereafter MV) metric instead 
of Friedmann-Robertson-Walker (FRW) metric. MV metric
approximates a spherical mass embedded in an asymptotically FRW spacetime. ``MV
mass" represents the
degree to which the void is not compensated by the mass of the shell.
They demonstrated the history of the shell in the flat MV spacetime,
showing that the negative MV mass acts to accelerate the shell expansion.

In this paper we extend the previous work to clarify the following points. 

(1) Peculiar velocities of uncompensated void, which is characterized 
by negative MV mass.
Although \citet{hai93} discussed the dynamics of voids with non-zero 
MV mass, the effect of MV mass on the peculiar velocity is not clear.
It is important to see it
because there is no evidence that actual voids have the shells with
compensated mass. For example, if voids originate from primordial
bubbles that are nucleated in a phase transition during inflation
(see, e.g. \citet{am96}), it is unlikely that voids have compensated shells.

(2) The relation between $\om$ and peculiar velocities was derived in 
LPT \citep{pee76}.  It is important to see how precisely the formula applies to 
nonlinear relativistic voids. We address this question, using the 
results obtained in (1) as well as the previous results \citep{SMS93}.

(3) As we mentioned above, \citet{pim86,pim88} arrived at the surprising
conclusion that
the effect of CMB radiation is significant. If it is true, we should 
take it into account seriously when we constrain the cosmological 
parameters from the observation of bulk flow. Therefore,
their result deserves closer examination.

This paper is organized as follows. In section 2, we present the relativistic 
equations of motion for a thin shell in MV spacetime, which will be solved 
later numerically. In section 3, we investigate how the
peculiar velocity changes due to MV mass for uncompensated voids. 
In section 4, we compare the results for the present model and those in the
LPT.
In section 5, we examine the effect of the CMB radiation on the void evolution.
These results are summarized in section 6.

%%%%%%%%%%%%%%%%%%%%%%%%%%%%%%%%%%%%%%%
\section{Basic Equations}
%%%%%%%%%%%%%%%%%%%%%%%%%%%%%%%%%%%%%%%
\subsection{McVittie Spacetime}

The spacetime described by MV metric has several useful properties:
\begin{itemize}
\item The near-field limit is Schwarzschild, in isotropic coordinates;
\item The far-field is a FRW spacetime;
\item The energy-momentum tensor has a perfect-fluid form.
\end{itemize}
It is an exact embedding of the Schwarzschild metric into the FRW metric.

The line element is 
\beq
ds^2 = -\left({1-h\over1+h}\right)^2dt^2
+a^2(t)(1+h)^4\bigl\{d\chi^2+f^2(\chi)(d\theta^2+\sin^2\theta d\varphi^2)
\bigr\}, \label{mv} 
\eeq
where
\beq
f ( \chi ) \equiv \left\{ \begin{array}{lll}
\sin \chi & (k=+1)& \\
\chi & (k=0)& \\ 
\sinh \chi & (k=-1)&
\end{array} \right.
 ~~~ {\rm and} ~~~
h\equiv{m\over4a(t)f(\chi/2)}
\eeq
The Einstein equations yield
\bea
{8\pi G\rho\over3}&=&H^2+{k\over a^2(1+h)^5},\label{rho}\\
8\pi Gp&=&{1\over1-h}\left[
-{2(1+h)\over a}{d^2a\over dt^2}-(1-5h)H^2-{k\over a^2(1+h)^5}\right],\label{p}
\eea
where $\rho$ and $p$ are the energy density and the pressure, 
respectively, which are inhomogeneous in general. $H\equiv (da/dt)/a$ is the
Hubble parameter at 
$\chi\rightarrow\infty$. $m$ is a constant and called MV mass. The
$h\rightarrow0$ limit is clearly FRW, while
Schwarzschild solution is recovered by $a\rightarrow1$. If $k=0$ 
or $-1$, the scale factor $a(t)$ always takes the Friedmann solution.

To see the meaning of MV mass, let us calculate the local gravitational 
mass defined by \citet{mis64}:
\beq
M\equiv{R\over2G}(1-g^{\mu\nu}\partial\lmu R\partial _{\nu} R)
~~ {\rm with} ~~~ R\equiv\sqrt{g_{\theta\theta}}.
\eeq
For each $k$ we find
\beq
M(R)={4\pi\over3}\rho R^3+m y(\chi) ~~~ {\rm with} ~~~ 
y(\chi)=\left\{ \begin{array}{ll}
\cos^5(\chi/2) & (k=+1) \\
1 & (k=0) \\ 
\cosh^5(\chi/2) & (k=-1)
\end{array} \right.
\eeq
As long as the void's size is much smaller than the horizon scale,
$\chi\ll 1$ and $y(\chi)\approx 1$ for any background model. In this
limit, we may 
therefore interpret the MV mass as approximately the Misner-Sharp mass 
minus the background mass $(4\pi/3)\rho R^3$.

For a thorough discussion of the
McVittie metric, its history, and its place among inhomogeneous models,
see \citet{kra}.

\subsection{Junction Conditions}

Let us derive the equations of motion for a spherical shell around a void, by
developing 
the thin-shell formalism of \citet{isr66}. The basic equations for the shell in 
the flat MV spacetime were given by \citet{hai93}.
Here we rewrite the equations as in a simpler form, which also describe the
shell in a
closed or open background, by extending the equations of Sakai, 
Maeda, \& Sato (1993). Because we are interested only in 
the effect of the outer MV mass, we assume the inside region to be homogeneous
throughout the paper.

Let a time-like hypersurface $\Sigma$, which denotes the world-hypersurface of
a spherical shell, divide a spacetime 
into two regions, $V^+$ (outside) and $V^-$ (inside). We define a unit
space-like vector
$N\lmu$, which is orthogonal to $\Sigma$ and pointing from $V^-$ to $V^+$. It
is convenient to
introduce a Gaussian normal coordinate system $( n, x^i )$ 
\footnote{In this paper, Greek letters run from 0 to 3, while Latin letters run
in 0, 2, 3.}
in such a way that the hypersurface of $n=0$ corresponds to $\Sigma$. From the
assumption that a shell is infinitely
thin, the surface energy-momentum tensor of the shell is defined as
\beq
S_{\mu\nu} \equiv \lim_{\epsilon \rightarrow 0} \int ^{\epsilon}_{-\epsilon}
T_{\mu\nu} dn .
\eeq

Using the extrinsic curvature tensor of the hypersurface of the shell, $K\ijl
\equiv N_{i;j}$, and the
Einstein equations, we can write down the jump conditions on the shell 
as \citep{ber87}
\beq
[K\ijl] ^{\pm} = - 8\pi G \Bigl( S\ijl- \frac12 h\ijl{\rm Tr}S \Bigr),
\label{jc1}
\eeq\beq
-{S_{i}^{j}}_{|j} = [T^{n}_{i}]^{\pm}, \label{jc2}
\eeq\beq
{{K\ij}^+ + {K\ij}^- \over 2} S_{i}^{j} = [T^{n}_{n}]^{\pm}, \label{jc3}
\eeq
where $h\ijl$ denotes the three metric of $\Sigma$, and $~|$ denotes
the covariant derivative with respect to $h\ijl$. We have denoted the
value of any field variable $\Psi$ on $\Sigma$ on the side of
$V^{\pm}$ by $\Psi^{\pm}$ and defined a bracket as 
$[\Psi]^{\pm} \equiv \Psi^+ - \Psi^- $. Eliminating $K\ijl^{~-}$ from 
equations (\ref{jc1}) and (\ref{jc3}), we can derive the equation 
\citep{ber87}:
\beq
{K\ij}^+ S^j_i + 4\pi G \Bigl\{ S\ij S^j_i - \frac12 ({\rm Tr} S) ^2 \Bigr\} =
[T^n_n]^{\pm}. \label{jc4}
\eeq

If the outer region is homogeneous, equation (\ref{jc4}) leads to a simple
expression of the basic equation, 
because it does not contain the metric in $V^-$ 
\cite{sak93a,sak93b,SMS93}. On the other hand, if we assume the inside region
to be
homogeneous, which is the case here, it is easier to solve the
equation obtained by eliminating $K\ijl^{~+}$:
\beq
{K\ij}^- S^j_i - 4\pi G \Bigl\{ S\ij S^j_i - \frac12 ({\rm Tr} S) ^2 \Bigr\} =
[T^n_n]^{\pm}. \label{jc5}
\eeq

The line elements in $V^+$ and in $V^-$ are described by
\beq
ds^2 = -\left({1-h\over1+h}\right)^2dt^2_+
+a_+^2(t_+)(1+h)^4\bigl\{d\chi_+^2+f_+^2(\chi_+)(d\theta^2+\sin^2\theta
d\varphi^2) \bigr\}, 
\eeq \beq
ds^2 = -dt^2_-+a_-^2(t_-)\bigl\{d\chi_-^2+f_-^2(\chi_-)(d\theta^2+\sin^2\theta
d\varphi^2), \bigr\}, 
\eeq
The direct calculation of ${K\ij}^{-}$ yields \citep{sak93a}
\beq
{K^{\theta}_{\theta}}^{-} = { \gamma_{-} ( f'_{-} + v_{-}H_{-}R) \over R}, ~~ 
{K^{\tau}_{\tau}}^{-} = \gamma_{-}^3 {dv_{-}\over dt_{-}} +\gamma_{-}
v_{-}H_{-}, \label{kij}
\eeq
where the circumference radius of the shell $R$, the peculiar velocity
of the shell $v_-$, and its Lorentz factor $\gamma_-$ are defined as
\beq
R \equiv a_+f_+ = a_-f_-,~~ v_{-} \equiv a_{-}{d\chi_{-}\over dt_{-}}, ~~ {\rm
and} ~~
\gamma_{-} \equiv {1\over\sqrt{1-v_{-}^2}}.
\eeq
Similarly, $v_+$ and $\gamma_+$ in the MV spacetime are defined as
\beq
v_{+} \equiv {a_+(1+h)^3\over1-h}{d\chi_{+}\over dt_{+}} ~~ {\rm and} ~~
\gamma_{+} \equiv {1\over\sqrt{1-v_{+}^2}}.
\eeq

As energy-momentum tensors, we consider perfect fluid on $\Sigma$ and in
$V^{\pm}$, i.e., 
\beq
S_{\mu\nu} = (\sigma+\varpi) v\lmu^{\pm} v_{\nu}^{\pm}+\varpi h_{\mu\nu},
\label{smn}
\eeq \beq
T_{\mu\nu}^{~\pm} = (\rho^{\pm}+p^{\pm})u\lmu^{\pm}u_{\nu}u^{\pm}+p g_{\mu\nu},
\label{tmn}
\eeq
where $\sigma$, $\varpi$, $v\lmu$, and $u\lmu$ are the surface
density, the surface pressure, the four velocity of the shell, and the
four velocity of the background fluid, respectively. In the Gaussian normal
coordinate system, we have
\beq
{T^n_n}^{\pm} = \gamma^2(v^2\rho+p)|^{\pm},~~ 
{T^n_{\tau}}^{\pm} = \gamma^2v(\rho+p)|^{\pm}. \label{tnt}
\eeq

Now, with the help of equations (\ref{kij}), (\ref{smn}), and (\ref{tnt}), we 
can write down equations (\ref{jc2}) and (\ref{jc5}) explicitly as
\bea
\gamma_-{d\sigma\over dt_-} &=& -2\gamma_-{dR\over dt_-}{\sigma+\varpi\over R}
+[\gamma v(\rho+p)]^{\pm}, \label{eom1}\\
\gamma_-^3{dv_-\over dt_-}&=&
-\gamma_-\left\{\left(1-2{\varpi\over\sigma}\right)v_-H_-
-{2f'_-\over R}{\varpi\over\sigma}\right\} -2\pi G(\sigma+4\varpi)
- {[\gamma^2(v^2 \rho+p)]^{\pm}\over\sigma}. \label{eom2}
\eea
The relation between $dR / dt_-$ and $v_-$ is given by
\beq
{dR \over dt_-} = f'_- v_- + H_- R . \label{eom3}
\eeq
Further, the conditions of the continuity of the metric,
\beq
d\tau^2 =\left({1-h\over1+h}\right)^2dt_+^2-a_+^2(t_+)(1+h)^4d\chi_+^2
= dt_-^2-a_-^2(t_-)d\chi_-^2,
\eeq \beq
{dR \over d\tau} = {d \over d\tau} \left\{(1+h)^2a_+f_+\right\} = {d \over
d\tau} (a_-f_-),
\eeq
reduce to 
\beq
{dt_+\over dt_-} = {1+h\over1-h}{\gamma_+\over\gamma_-}, \label{eom4}
\eeq \beq
\gamma_+ \left\{\left(f'_++{2h'f_+\over1+h}\right)v_+ + H_+R\right\} =
\gamma_- (f'_-v_- + H_-R). 
\label{eom5}
\eeq
The equations of motion for the shell are determined by equations (\ref{eom1}),
(\ref{eom2}), and
(\ref{eom3}). We use the above supplementary equations (\ref{eom4})
and (\ref{eom5}) to give $t_+$ and $v_+$, respectively; they are used
at the initial time as well as at each step of time evolution. 

The angular component of the jump condition (\ref{jc1}),
\beq
\gamma_+\left(f'_++{2h'f_+\over1+h}+v_+H_+R\right)-\gamma_- (f'_- +
v_-H_-R) = -4\pi G\sigma R, \label{eom6}
\eeq
gives a constraint for the relation between the surface density $\sigma$ and MV
mass $m$. We use
it for giving initial data as well as for checking numerical errors of 
integration. For integration we adopt the 4th order Runge-Kutta
method. Throughout the analysis we did not encounter any numerical
problem: the relative errors of equation (\ref{eom6}) were always less than 
$10^{-13}$. 

The equations of motion presented here and by \citet{SMS93} have
several advantages compared with the equations derived in other
papers. First, the expression is much simpler. Secondly, there is no sign
ambiguity in the
relation between $t_+$ and $t_-$, (\ref{eom4}), contrary to the comment
by Pim \& Lake (1986). Thirdly, our expression for the extrinsic curvature
$K^{\theta}_{\theta}$, (\ref{kij}), can take  both positive and negative values
without ambiguity. Although
our equations make numerical integration easier, they may not be so convenient 
for analytic arguments.

%%%%%%%%%%%%%%%%%%%%%%%%%%%%%%%%%%%%%%%
\section {Peculiar Velocities of Uncompensated Voids}
%%%%%%%%%%%%%%%%%%%%%%%%%%%%%%%%%%%%%%%

Here we consider only dust as matter fluid:
\beq
p^{\pm}_{\infty}=0 ~~ {\rm and} ~~ \varpi=0,
\eeq
where the subscript $\infty$ denotes quantities at $\chi_+\rightarrow\infty$.
A compensated void simply means $m=0$, i.e., the background is 
described by the FRW metric. On the other hand, an uncompensated void is 
characterized by negative MV mass. Here we fix the value of $m$ by 
supposing no shell ($\sigma_i=0$) at the initial time $t_i$.
The initial time $t_i$ is a free parameter, which is determined by the 
structure formation  model; in the following we set $t_i$ as the 
decoupling time , i.e., $z_i=1000$.
The remaining initial parameters are fixed as follows:
$$
v^+_i=0,~~ R_iH^+_i=0.1,~~ H^+_{i}=H^-_{i},~~ \rho^-_i=0,
$$ \beq
\Omega_i \equiv {8\pi G\rho^+_i\over 3H_i^2}=1 ~~ {\rm or} ~~ 0.98.
\eeq

Figure 1(a) shows the motion of the shell in terms of the comoving 
coordinate $\chi$. As shown by Haines \& Harris (1993),
negative MV mass pushes the shell faster. Although Figure 1(a) 
indicates that the effect of MV mass looks quite large, the 
behavior of $\chi$ is not observable. What we can observe is the radius 
and velocity of the shell at the present epoch.
We thus plot the peculiar velocity normalized by the Hubble expansion:
\beq\label{tv}
\tv \equiv {v\over HR}.
\eeq
in Figure 1(b). The asymptotic behavior is determined by $\Omega_0$, 
independent of MV mass. Figure 2 reports the relation between $\tv_0$ and 
$\Omega_0$, which confirms that
$\tv_0$ does not depend on whether the void is compensated or not.

%%%%%%%%%%%%%%%%%%%%%%%%%%%%%%%%%%%%%%%
\section{Comparison with Linear Perturbation Theory}
%%%%%%%%%%%%%%%%%%%%%%%%%%%%%%%%%%%%%%%

In this section we discuss the relation between $\om$ and $\tv_0$,
comparing the results in the relativistic void model and those in LPT.

The peculiar velocity {\boldmath$v$} for general density fluctuations in LPT is
\citep{pee76}
\beq\label{LPT}
\mbox{\boldmath$v$}={2F\mbox{\boldmath$g$}\over 3H\Omega} ~~~ {\rm with} ~~~
F\approx\Omega^{0.6},
\eeq
where {\boldmath$g$} is the peculiar gravitational acceleration. For a 
spherically symmetric system, the gravitational acceleration is given by
\beq\label{g}
g(R)=-{G\delta M(R)\over R^2},
\eeq
where $\delta M(R)$ is the difference between the mass within a sphere 
and the unperturbed mass within the sphere with the same radius $R$.

For the void model, $\delta M(R)$ depends on whether we measure it just inside
the shell ($R=R_-$) or just
outside it ($R=R_+$):
\beq
\delta M(R_-)=-{4\pi\over3}(\rho^+-\rho^-)R^3, ~~~
\delta M(R_+)=-{4\pi\over3}(\rho^+-\rho^-)R^3+4\pi\sigma R^2.
\eeq
It is therefore reasonable to define the mass difference as the average:
\beq\label{dM}
\delta M(R)\equiv{\delta M(R_+)+\delta M(R_-)\over2}
=-{4\pi\over3}(\rho^+-\rho^-)R^3+2\pi\sigma R^2.
\eeq
On the other hand, one of the junction conditions (\ref{eom6}) can be rewritten
as
$$
\varepsilon^+\sqrt{1+\left({dR\over d\tau}\right)^2-{8\pi
G\rho^+\over3}R^2-{2Gmy\over R}}-
\varepsilon^-\sqrt{1+\left({dR\over d\tau}\right)^2-{8\pi G\rho^-\over3}R^2}
$$
\beq\label{eom62}
=-4\pi G\sigma R,
\eeq
where $\varepsilon^{\pm}\equiv{\rm sign}{K^{\theta}_{\theta}}^{\pm}$. 
In the Newtonian approximation, $(dR/d\tau)^2\ll1$, $y(\chi)\approx1$, 
and $\varepsilon^{\pm}=+1$, equation (\ref{eom62}) reduces to mass
conservation:
\beq\label{eom63}
m+{4\pi\over3}(\rho^+-\rho^-)R^3=4\pi\sigma R^2.
\eeq
From equations.(\ref{tv}), (\ref{LPT}), (\ref{g}), (\ref{dM}), and
(\ref{eom63}), we obtain
\beq\label{tv2}
\tv={\Omega^{0.6}\over6}\left(1-{\rho^-\over\rho^+}-{m\over4\pi\rho^+R^3/3}
\right).
\eeq

For compensated voids ($m=0$), equation (\ref{tv2}) reduces to a simple 
expression:
\beq
\tv={\Omega^{0.6}\over6}\left(1-{\rho^-\over\rho^+}\right).
\eeq
For uncompensated voids ($\sigma_i=0$), on the other hand, 
equations (\ref{tv2}), (\ref{eom63}) and $\rho^+a_+^3=$const. read
\beq\label{tv3}
\tv={\Omega^{0.6}\over6}\left(1-{\rho^-\over\rho^+}\right)\left\{1+\left({r^+_{i
}\over r^+}\right)^3\right\}.
\eeq

Figure 3 shows plots of $\tv_0$ v.s. $\Omega_0$ for compensated 
voids, where the subscript 0 denotes quantities at the present. (The details 
of the analysis for the homogeneous background were given by \citet{SMS93}.
In the linear case (a), our numerical result is in good accordance with the
result in LPT. Even in the nonlinear case (b), the difference between 
the two results is relatively small (up to 10\%). 

Let us turn to the case of uncompensated voids. Obviously the term
$(r^+_{i}/r^+)^3$ in equation (\ref{tv3}) represents the effect of MV mass: as
the comoving radius $r$ increases, the effect of MV mass decreases. 
This argument explains the result that the eventual behavior of $\tv$ 
does not depend on MV mass, as shown in Figure 1(b).

%%%%%%%%%%%%%%%%%%%%%%%%%%%%%%%%%%%%%%%
\section {Effect of CMB radiation within a void}
%%%%%%%%%%%%%%%%%%%%%%%%%%%%%%%%%%%%%%%

As we mentioned in the introduction, Pim \& Lake (1986,1988) showed 
that, if we include CMB radiation, the shell expands much 
faster than that in the absence of radiation, and that its asymptotic
behavior is $R\propto t$ even in the flat universe. Here we re-examine 
the effect of radiation in the flat FRW background: $m=0$ and $k^+=0$.

First, let us reproduce the results of Pim \& Lake (1986). As background
matter,
a mixture of dust ($\rho_d^+$) and blackbody radiation ($\rho_r^+$) 
is considered:
\beq
\rho^+=\rho_d^++\rho_r^+=\rho_{cr}\equiv{3H_+^2\over8\pi G},~~~
p^+={\rho_r^+\over 3}
\eeq
Setting the present temperature Hubble parameter as $T_0=2.7$K and 
$H^+_0=100/$kms/Mpc and the using the relation,
\beq
\rho_r^+={8\pi^5\over15}{k_B^4T^4\over h^3c^3},
\eeq 
the background model is completely fixed. The interior is assumed to
be the flat FRW spacetime with radiation only, of which abundance 
($\rho_r^-$) is characterized by a parameter,
\beq
\alpha\equiv\left({\rho_r^-\over\rho^+}\right)_i.
\eeq
For matter fluid on the shell, they assume that the equation of state has a
form,
$\varpi=\epsilon\sigma$. In one of their calculations, the initial parameters 
are fixed as follows:
$$
z_i=1000,~~ v^+_i=0.1,~~ R_iH^+_i=0.1,~~ \epsilon=0,~~ k_-=0,
$$ \beq
\alpha=10^{-1},~10^{-2},~10^{-3},~10^{-4},~{\rm or} ~10^{-5}.
\eeq
Integrating the equations of motion in section 2, we obtain the result
in Figure 4(a), which is the reproduction of Figure 5 of Pim \& Lake (1986).

This figure tells us that, no matter how little radiation exists, it affects 
the shell's motion significantly. Because this result 
was surprising to us, we examine their analysis. As a result, we 
find that their assumption of $k^{-}=0$ was inappropriate for the 
following reason. If
$k^{\pm}=0$ and $\rho^+>\rho^-$, the Friedmann equation reads $H^+>H^-$. As
Sato (1982) and Sato \& Maeda (1983) argued, however, a thin shell is 
formed by compression of matter like a 
snow-plow mechanism when the inner expansion is faster than the 
background expansion, i.e., $H^-<H^+$. Therefore, the assumption of 
$k^{\pm}=0$ is inconsistent with the thin shell description.
If we still used the thin-shell equations for the case where the shell expands 
faster than the interior matter 
fluid, a part of the shell mass would be forced to ``evaporate'' 
so as to keep the 
inside homogeneous, i.e., the shell would emit mass and accelerate, 
which seems unphysical.
This explains the odd behavior in Figure 4(a).

Here we re-analyze voids with CMB radiation. Although 
the exact value of $H^-_i/H^+_i$ cannot be determined without 
knowing the formation process, the consistency with thin shell 
requires $H^-\ge H^+$. Because we still assume the background universe 
to be flat, the inside region should be open. Adopting $H^+_i=H^-_i$ 
instead of $k^{-}=0$ and leaving the other conditions unchanged, we solve the
equation of 
motion. The result is reported in Figure 4(b), showing that the effect of
radiation is
much smaller.

We should note, however, that it is not so fruitful to investigate 
further details of the motion of voids including radiation in this approach. 
Because we do not know the physical process of radiation around the shell, the 
equation of state ($\epsilon$) is not determined; furthermore, even
the validity of the thin-shell approximation is not clear.
What we can conclude is that, under the condition that the thin-shell 
approximation is valid, the effect of CMB radiation on void expansion is 
negligible.

%%%%%%%%%%%%%%%%%%%%%%%%%%%%%%%%%%%%%%%
\section {Summary}
%%%%%%%%%%%%%%%%%%%%%%%%%%%%%%%%%%%%%%%

As a model of nonlinear structure, we have considered a relativistic
void in the expanding universe, and discussed peculiar velocities.

(1) In order to investigate the dynamics of shells with uncompensated 
mass, we have adopted McVittie spacetime as a background universe. 
Although the motion itself is quite different from that of a 
compensated void, as shown by Haines \& Harris (1993), 
the present peculiar velocities are unaffected by MV mass.

(2) We discuss the relation between $\om$ and peculiar velocities, 
comparing the results in the present model with those 
in the linear perturbation theory. For nonlinear voids, the quantitative
difference between 
these two results is up to 10\%, which is relatively small.

(3) Because Pim \& Lake (1986,1988) arrived at the surprising conclusion that
the effect of a small amount of CMB radiation is significant, we have 
re-examined it. We have shown that their
results are due to inappropriate initial conditions. With modified 
initial conditions, the effect of radiation turns out to be negligible. 

Although we have investigated only specific models of nonlinear 
structure, our results (1)-(3), as a whole, indicate that the formula 
for peculiar velocities in the linear 
perturbation theory can apply approximately to nonlinear voids.

\acknowledgements

Numerical Computation of this work was carried out at the Yukawa Institute
Computer Facility. 
N.S. was supported by JSPS Research Fellowships for Young Scientists,
No.9702603.

%%%%%%%%%%%%%%%%%%%%%%%%%%%%%%%%%%%%%%%%%%%%%%%%%%%%%%%%%%%%%%%%%%%%%%%%

\end{document}